\documentclass{ws-procs961x669}

\begin{document}
\title{Dark gravitomagnetism with LISA and gravitational
waves space detectors}

\author{A. Tartaglia$^*$ }

\address{OATo, INAF and DISAT, Politecnico\\
Corso Duca degli Abruzzi 24, 10129 Torino, Italy\\
$^*$E-mail: angelo.tartaglia@inaf.it\\
www.polito.it}

\author{M. Bassan and G. Pucacco}

\address{Dip. Fisica, Tor Vergata University and INFN/Roma2, Via della Ricerca Scientifica 1\\
00133 Rome, Italy\\
E-mail: bassan@roma2.infn.it}

\author{V. Ferroni and D. Vetrugno}

\address{Dip. Fisica, Trento University and TIFPA, Via Sommarive 14\\
38123 Povo TN, Italy\\
E-mail: valerio.ferroni@unitn.it}

\begin{abstract}
We present here the proposal to use the LISA interferometer for detecting the gravitomagnetic field due to the rotation of the Milky Way, including the contribution given by the dark matter halo. The galactic signal would be superposed to the gravitomagnetic field of the Sun. The technique to be used is based on the asymmetric propagation of light along the closed contour of the space interferometer (Sagnac-like approach). Both principle and practical aspects of the proposed experiment are discussed. The strategy for disentangling the sought for signal from the kinematic terms due to proper rotation and orbital motion is based on the time modulation of the time of flight asymmetry. Such modulation will be originated by the annual oscillation of the plane of the interferometer with respect to the galactic plane. Also the effect of the gravitomagnetic field on the polarization of the electromagnetic signals is presented as an in principle detectable phenomenon.
\end{abstract}


\bodymatter

\section{Introduction}\label{aba:sec1}
Among the possible space-times consistent with the Einstein equations and with various matter distributions, a special interest  deserve those endowed with a chiral symmetry about the time axis of a given observer. The chiral symmetry may sometime appear as a simple local coordinates artifact, in which case the symmetry disappears with an appropriate choice of the reference frame. Clearly,  the most interesting situations are when the symmetry is indeed in the curvature of space-time, which happens when the source of gravity is a spinning mass distribution. In terms of the metric tensor, the presence of the symmetry we are interested in is manifested by non-null time-space off-diagonal terms. It is of course always possible to choose a local reference frame where all off-diagonal terms are brought to zero, but this is in general not possible \textit{globally}, when the source of gravity is spinning. In weak field conditions (the most common situation in the universe) the symmetry we are considering appears in the form of a \textit{gravito-magnetic} field. The name derives from the fact that in weak field approximation the Einstein equations assume a form quite similar to Maxwell's equations of electromagnetism (e.m.), so that the interaction between moving masses (just as moving charges in e.m.) may be described in terms of  forces due to two vector fields: the \textit{gravito-electric} (analogous to the electric field, but always attractive), and the  \textit{gravito-magnetic} field (analogous to the magnetic field of classical e.m.). This formalism is well known and we refer to the  vast literature and to basic texts of general relativity (GR): see for example Ref. [\citenum{gravitation}].

It is, however, worth remarking that the gravito-magnetic force is usually much weaker than the gravito-electric one, making thus hard to identify measurable effects. So far, physically relevant phenomena associated with the aforementioned symmetry are expected, upstream of any approximation, in strong field conditions such as near Kerr black holes or, when the gravito-electromagnetic approximation is applied, in the behaviour of peculiar systems where gravity is still strong enough to allow the gravito-magnetic component to emerge in an observable way. The latter is the case of the double pulsar, whose internal dynamics, readable from the signals received by our radiotelescopes, indirectly hints to the gravito-magnetic (GM) interactions between the spins and the orbital motions of the two stars in the pair.\cite{Breton}

In the Solar System, the effects due to the angular momenta of the Sun and of the planets are extremely weak. Until now the only measured effects are those of the terrestrial gravito-magnetic field. They have been observed monitoring the motion of the moon along its orbit by Laser ranging\cite{lunar} and doing the same with the LAGEOS and LAGEOS2 satellites:\cite{Ciufolini} what was found was the Lense-Thirring drag \cite{lense} on their motion. Direct evidence has been obtained by the dedicated Gravity Probe B experiment which  measured the induced precession on gyroscopes carried on a circumterrestrial satellite in polar orbit.\cite{everitt} Laser ranging is also the main tool used to analyze the orbit of the dedicated LARES mission, which is producing the best accuracy so far for the GM field of the Earth.\cite{Larase2020}
On the surface of our planet it is expected to measure the GM field using ring lasers.\cite{bosi,divirgilio,tartaglia2017}

A way to compensate for the weakness of the sought effects is to use very large sensors and measuring devices, that could not be hosted on a single spacecraft or in a laboratory on Earth. This is one of the reasons pushing toward peculiar configurations of a plurality of measuring devices in space, either around the Earth\cite{ruggiero} or at the scale of the inner solar system.\cite{tartaglia2021}

An extremely interesting opportunity is the LISA interferometer \cite{LISA}, designed and under development for detecting gravitational waves. What we propose here is to use LISA, which, as we shall see, is already fit for the purpose, also to reveal the GM field in which the device is immersed. The interesting sources that could be studied are both the solar angular momentum and the angular momentum of the Milky Way (or, to say better, the part of it which is relevant for the solar system). A further reason of interest for the galactic angular momentum is that it reasonably depends both on the visible and on the dark component of our galaxy. So far, dark matter (DM), which is an important ingredient of the cosmic cocktail, is postulated only on the basis of its direct gravitational (gravito-electric) effect. However,  if DM exists and we trust GR, we may expect also phenomena related to the proper rotation of DM distributions.  It is currently accepted that most galaxies (including ours) are immersed in huge DM halos; if so, it is also reasonable to expect such halos to rotate together with the visible part of the galaxy: in fact, if dark and visible interact gravitationally, any little localized inhomogeneity on one side or the other acts as a hook that drags or brakes the other component so that, after some remote transitory, both end up revolving together. The consequence is that DM should represent a relevant contribution to the angular momentum of the whole, then to the GM component of the gravitational interaction.

In what follows we outline how LISA could be used to measure both the angular momentum of the Sun and of the Milky Way. Reliable estimates of  the angular momentum of the visible Milky Way exist:  by difference, we could thus measure the galactic Dark content.

\section{Internal space-time of the Milky Way}

Globally and in the first instance we can attribute an axial symmetry to the distribution of matter of the Milky Way and imagine also that it is in uniform rotation condition (stationary space-time).
If so the typical line element may be written as follows:

\begin{equation}
ds^2=U(r,z) c^2 dt^2-2N(r,z)r d\phi c dt-W(r,z)r^2d\phi^2-Q(r,z)(dr^2+dz^2)\label{eq1}
\end{equation}

Cylindrical space coordinates have been used. An additional symmetry we assume is reflection symmetry about the galactic plane. $U,N,W,Q$ are dimensionless functions of $r$ and $z$ only. They are everywhere regular (except possibly in the origin). The only constraint expressing the reflection symmetry is \begin{equation}
{\frac{\partial U}{\partial z}}\big|_{z=0}={\frac{\partial N}{\partial z}}\big|_{z=0}={\frac{\partial W}{\partial z}}\big|_{z=0}={\frac{\partial Q}{\partial z}}\big|_{z=0}=0
\end{equation}

Our interest is on the view point of an observer at rest with the Sun, so for the moment it is convenient to assume that the $r$ axis is corotating with the Milky Way. This assumption does not spoil the symmetry we have assumed, but implies that the $U,N,W,Q$ functions implicitly include also the effects of the choice of the observer (which in general treatments is assumed to be at rest with the center of the mass distribution and located at infinity, whereas here it is co-moving with the Sun).

We cannot a priori say more about the shape of the functions because we are considering an observer inside the mass distribution (visible or dark it may be) and we do not know exact solutions of Einstein's equations within a rotating distribution of matter. Nonetheless, if we succeed in detecting specific effects of the rotation of the Milky Way, we may also deduce relevant information on the way DM is distributed. This is because the non trivial parts of the elements of the metric tensor contain the mass density of the Milky Way and the rotation curve of the stars in it: since we know how the visible mass is distributed and the rotation curve of the galaxy (which we have assumed to be the same for DM also), from the evaluation of the total GM field we may deduce relevant information on the DM density distribution, which contributes to the effect.

\subsection{Galactic gravito-magnetism}

Line element (\ref{eq1}) is quite general, but we may equally well adopt the language and the images of the weak field, since that is indeed the case. We may also focus our attention on the situation on and near the galactic plane considering that the Sun is not far from it. This said, we read the ratio $N/U$ as the only non-zero component of a three-vector, which is the analog of the vector potential $\bar{A}$ of classical electromagnetism. \cite{tartaglia2005}

Under the above conditions, whenever the metric tensor is independent of time the equation of geodesic motion can be written in a form analogous to Lorentz's equation of e.m.
\begin{equation}
\frac{d^2\bar x}{dt^2}  = - c^2(\bar \nabla U - 2 \frac{\bar v}{c} \times \bar \nabla \times \bar A_g)
\end{equation}
$\bar v$ and $\bar x$ are the three-dimensional velocity and position of the test mass with respect to the observer.
The functions $U$ and $\bar A_g$ are given in terms of the elements of the metric tensor $g_{\mu\nu}$ by:

\begin{equation}
 U= g_{00};~~  A_{(g)i} = \frac {g_{0i}}{g_{00}}
 \end{equation}
 In our case it is
 \begin{equation}
 \bar{A}_g = (0,\frac{N}{U},0) ~~\textrm{or}~~ \bar{A}_g= \frac {N}{U} \hat \phi
\end{equation}
Here $\hat \phi$ is the unit transverse vector in the direction of rotations about the symmetry axis.
Our space basis for covariant vectors is $(dr,rd\phi,dz)$ so that all components of $\bar{A}_g$ are dimensionless and the field lines of the vector are circles centered on the space symmetry axis, contained in planes perpendicular to that axis.
The next step, formally recognizing the weak field approximation, is to calculate a \textit{gravito-magnetic field} $\bar{B}_g$ from $\bar{A}_g$ just in the same way we would do in classical electromagnetism:
\begin{equation}
\bar{B}_g = \bar{\nabla}\times \bar{A}_g=\frac{N}{U}(\frac{\partial_z{U}}{U}-\frac{\partial_z{N}}{N})\hat{r}+\frac{N}{U}(\frac{\partial_r{N}}{N}-\frac{\partial_r{U}}{U})\hat{z}
\label{bg}
\end{equation}
$\partial_r$ and $\partial_z$ are a shorthand notation for $\partial/\partial{r}$ and $\partial/\partial{z}$; each component of $\bar{B}_g$ has the dimension of the inverse of a length.

If we limit ourselves to the symmetry plane (the galactic plane) the $r$ component of the field vanishes and the vector turns out to be perpendicular to the plane, then parallel to the axis of the Milky Way.

\section{The propagation of light}

In the peculiar space-time we are considering, the propagation of light displays an interesting behaviour. Whenever some physical device (mirrors, waveguides or else) constrains light to move along a closed path, the time of flight (ToF) it takes to make a full turn is different depending on whether the travel is right- or respectively left-handed. The sense of the circulation is with respect to the field lines of $\bar{B}_g$.

Actually this asymmetry is not peculiar of light only. It holds true for any signal or voyager, provided its velocity is locally the same in both directions at any position along the path.\cite{ruggiero2015} The effect we are recalling is sometimes improperly called 'Sagnac effect', though the latter is an effect of special instead of general relativity and concerns the non inertial rotational motion of the observer rather than the curvature of space-time.

When considering the situation from a four-dimensional viewpoint, light leaves the source at a given event of the worldline of the observer, then comes back to the observer at a different point along its worldline: the path is of course open. When starting in the opposite direction, the light ray will meet again the observer at a different event of its worldline. Locally measuring the proper time ($\tau$) difference between the arrivals (the length of the intercepted interval of the observer's worldline) we deduce an information concerning the proper angular momentum of the source of gravity (possibly combined with rotational motion of the observer, if it is present).

The difference in proper time  can be expressed in terms of the components of the metric tensor. Using the standard notation of GR we have:\cite{kajari}
\begin{equation}
\Delta\tau=-\frac{2}{c}\sqrt{g_{00}}\oint\frac{g_{0i}}{g_{00}}dx^i \hskip12mm i= 1,2,3
\end{equation}
The square root of $g_{00}$ is evaluated at the position of the observer and accounts for the gravitational field there; it does not coincide with the $g_{00}$ under the integral sign, since the latter expresses the gravitational field along the integration path. Here $x^i$ is a generic space coordinate and the integration is performed along the space closed contour traveled by the light beam.

Restricting the description to the space-time with the symmetries described above, and using the typical notation of gravito-electromagnetism, the time delay becomes:
\begin{equation}
|\Delta\tau|=\frac{2}{c}\sqrt{U}\oint A_{gi} dx^i=\frac{2}{c}\sqrt{U}\int\bar{B}_g\cdot\hat{u}_ndS
\label{flux}
\end{equation}

The last term on the right is obtained applying the Stokes theorem of usual three-dimensional geometry and converts the line integral into a flux; $dS$ is the surface element of the area contoured by the closed path in space and $\hat{u}_n$ is the unit three-vector perpendicular to the given surface element.
\subsection{ Galactic Gravito-Magnetism in the Solar System}
Our purpose is to consider an experiment carried out at the scale of the inner solar system and indeed at the Earth's orbit. The galactic GM field $\bar{B}_g$ does in fact depend on the distance from the center of the Milky Way.
The Sun is approximately
$R_S\cong 2.35\times 10^{20}$m ($28,000$ light years) away from the center of the Milky Way.
If we think of an instrument revolving around the Sun in correspondence of the orbit of our planet, its distance from the center of the Milky Way changes at most by plus or minus $1$ astronomical unit (AU), i.e. $\Delta R_S\sim 3\times 10^{11}$ m. The relative fluctuation of the distance, under these conditions, would then be $\Delta R_S/R_S\sim 10^{-9}$.
 Even though we have no explicit expression for the functions appearing in line element (\ref{eq1}), just looking at what happens in electromagnetic analogies, we may reasonably expect $\Delta B_g/B_g$  to be of the same order of magnitude as $\Delta R_S/R_S$. If so, if we additionally assume such a relative change to be negligible for our purposes, it follows that in practice $\bar{B}_g$ of the Milky Way is a constant for our experiment.
Under this assumption eq. (\ref{flux})  simplifies to:
\begin{equation}
|\Delta\tau|\cong\frac{2}{c}\sqrt{U}B_g S \cos{\gamma}
\label{base}
\end{equation}
$S$ is now the total area enclosed in the path of light, assuming that the contour is contained in a plane; $\gamma$ is the angle between the perpendicular to that plane and the direction of the gravito-magnetic field of the Milky Way at the Sun. As far as the Sun may be considered to lay in the galactic plane, the direction of $\bar{B}_g$ coincides with that of the axis of the Milky Way.

If we are able to measure $|\Delta\tau|$, we may then deduce from eq. (\ref{base}) information on $U$ and in particular on $N$, since $U$, being related to the local ordinary gravitational potential, can also be obtained by other means.

\section{Space interferometers and LISA}
It follows from  previous considerations and from eq. (\ref{base}) that a good strategy to detect the presence of a gravito-magnetic field is to resort to the measurement of time asymmetries in the propagation of light; eqs. (\ref{flux}) and (\ref{base}) show that we need a vast area {$S$}  in order to enhance the sensitivity of the experiment. This is why appropriate configurations of receivers and transponders in space are particularly interesting.
For instance, it has been proposed\cite{ruggiero} to use constellations of satellites around the Earth (such as the Galileos), but even more appropriate and promising may be the opportunity to use LISA, an orbiting interferometer designed to detect gravitational waves around the mHz frequency band.

\begin{figure}
\begin{center}
\includegraphics[width=0.43\textwidth]{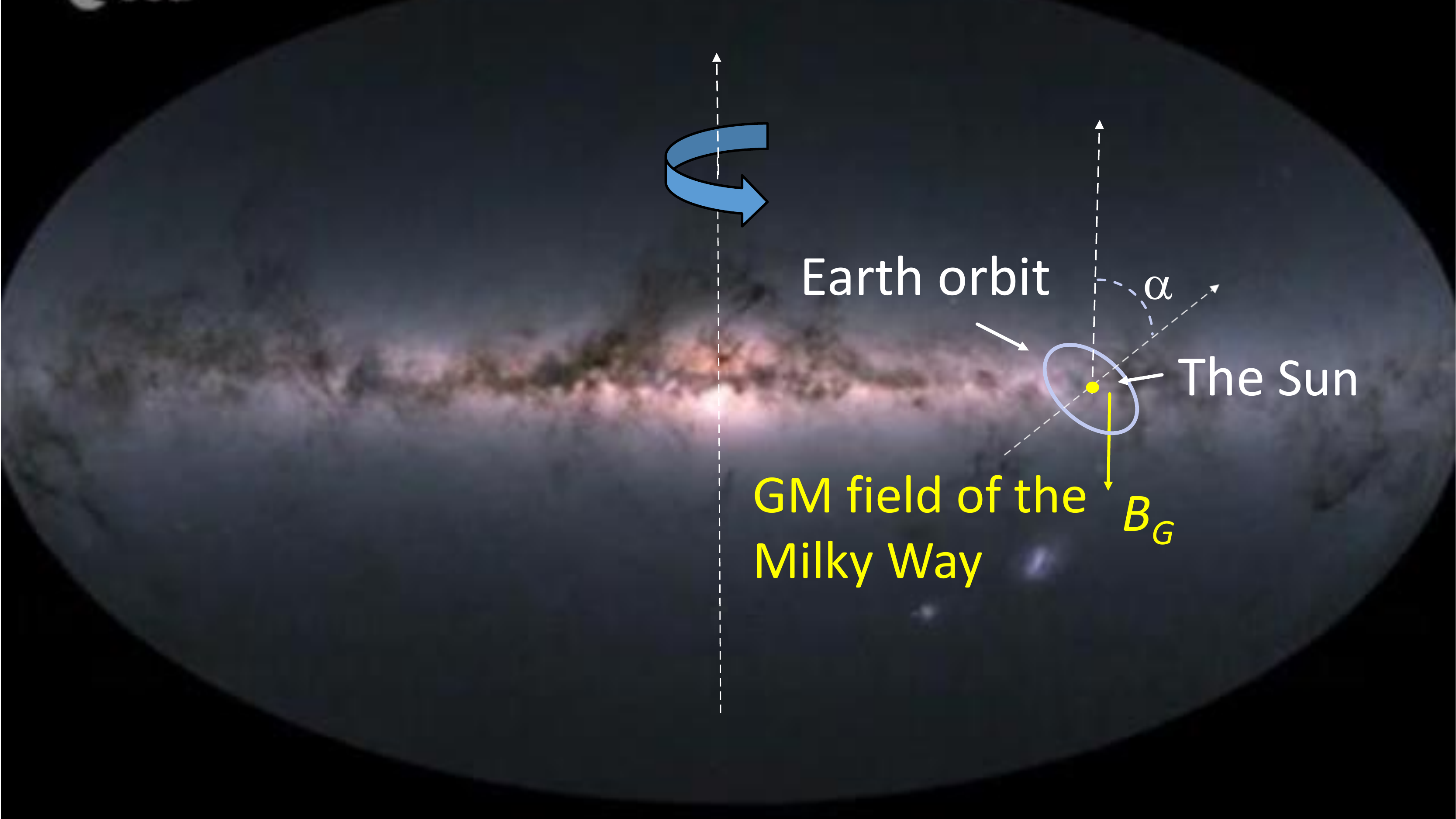}
\includegraphics[width=0.52\textwidth]{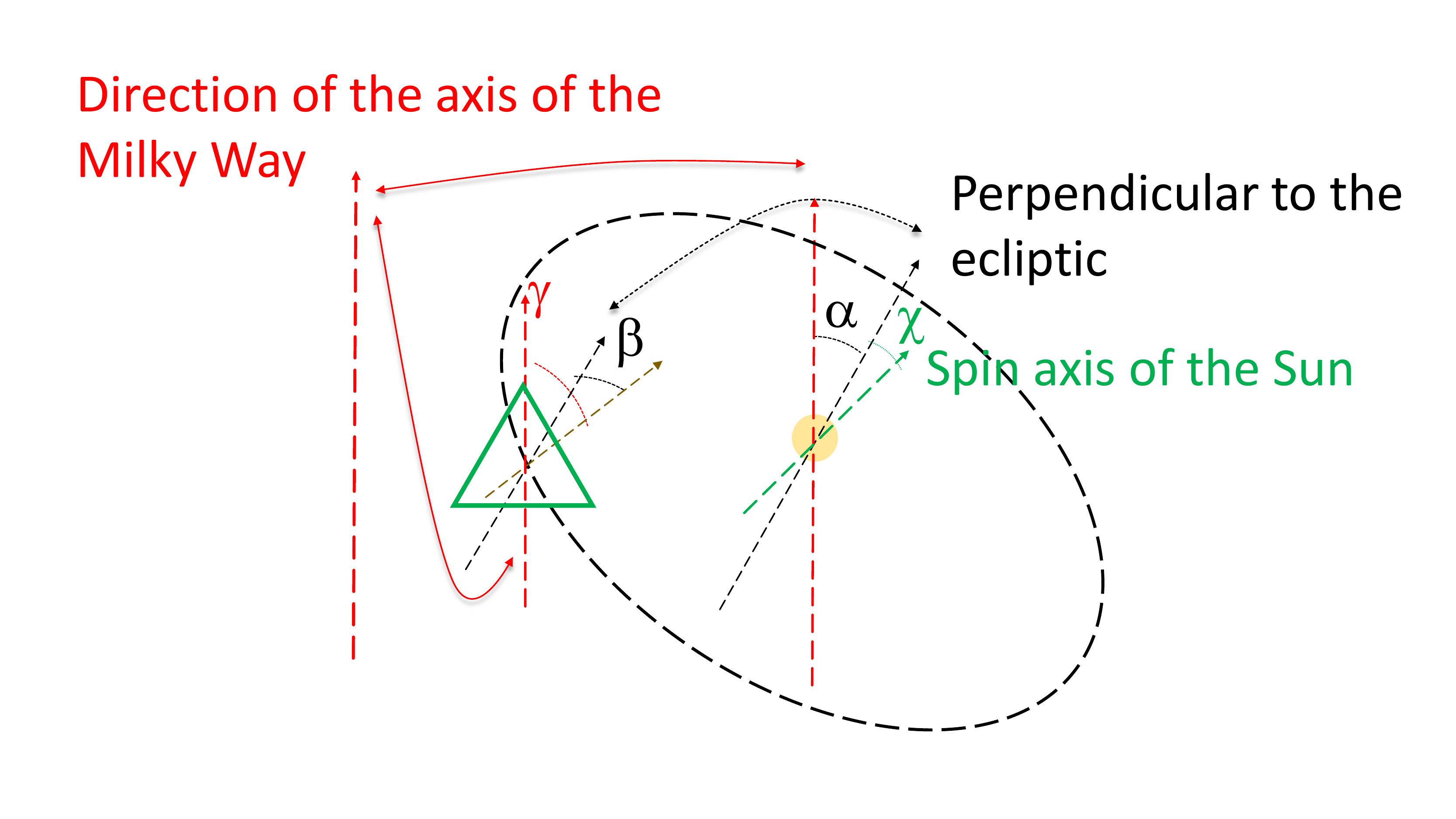}
 \caption{The LISA interferometer in the Milky Way.
 a) (left) Freely falling constellation in the terrestrial orbit. b) (right) Orientation of the triangular constellation with respect to the galaxy and to the ecliptic.}
\label{fig1}
\end{center}
\end{figure}

LISA will monitor the distance between pairs of free falling test masses (TM) at a distance $L = 2.5$ million km with a continuous  interferometric laser ranging scheme to detect strain distance variation due to gravitational waves. The TM pairs will be lodged in three spacecraft arranged to form a big equilateral triangular constellation.
So far we considered the observer as being co-moving with the Sun, but actually the plane of the LISA constellation will form an angle $\beta = 30^o$ with the ecliptic plane and that plane is at an angle $\alpha\simeq 60^o$ with respect to the galactic plane; the whole constellation will rotate around the Sun at a distance of 1 AU, lagging by $20^o$ the orbit of the Earth; it will also rotate on its plane, around the center of mass, with the same 1 year period. The geometry of such configuration is shown in Fig.(\ref{fig1}).

Keplerian dynamics causes, in the yearly rotation period, changes of up to $10^4$ km in armlength and $10$ m/s in spacecraft velocity: this phenomenon, called {\it flexing} of the arms,  forbids equal path interferometry. However, the use of post-processing techniques (Time Delay Interferometry: TDI)\cite{TDI,TDI1}  allows to synthesise equal arm interferometry to better than 1 ns.
Among TDI combinations there are several that mimic the output of a Sagnac interferometer \cite{TDI2,TDI3}, thus suppressing the GW signals.
By interfering electromagnetic signals  traveling in opposite directions along the LISA triangle, we could measure the ToF difference at one of the corners: Eq. (\ref{flux}) will then allow us to deduce the intensity of the gravito-magnetic flux through the interferometer.

In the  reference frame comoving with the spacecraft, $\bar{B}_g$ will contain various contributions including those due to the proper and orbital motion of the device (Sagnac effect). The components of interest would be the galactic contribution (both from visible and dark matter) and the gravito-magnetic field originating from the angular momentum of the Sun. In principle also the Earth and other planets would contribute, but it is easily verified that those components would indeed be much weaker than the galactic and solar ones.

\subsection{Gravito-magnetism from the Sun}
The GM field of the Sun, $\bar{B}_{g\odot}$, has a simple form, as long as our star is treated as a compact spherical spinning source. The configuration of the field is dipolar and in the solar equatorial plane, where $\bar{B}_{g\odot}$ is perpendicular to the plane, it is (net of kinematic contributions, i.e. for a non-rotating observer in a Sun-centered reference frame)\cite{bosi}

\begin{equation}
\bar{B}_{g\odot} = \frac{2G}{c^3 r^3}\bar{J}_\odot = 2.8 \cdot 10^{-28} \textrm{m} ^{-1}
\label{Sun}
\end{equation}
$G$ is Newton's constant and $\bar{J}_\odot$ is the angular momentum of the Sun; $r$ is the observer's distance from the center of the star.

In our approximation, we can take $U = 1-2\frac{GM_\odot}{c^2 r} \simeq 1$, and the ToF asymmetry, for the solar effect alone, would then be:

\begin{equation}
|\Delta\tau|_\odot \simeq
\frac{2}{c}\int |\bar{B}_{g\odot}| \cos{\eta}dS
\simeq
\frac{4G}{c^4}\int\frac{|\bar{J}_{\odot}|}{r^3}\cos{\eta}dS
\label{Sundelay}
\end{equation}

The flux is calculated over the LISA triangle:
actually, the area of the constellation is tilted with respect to the ecliptic, and this has two consequences: 1) part of the area is slightly above and part slightly below the ecliptic plane so that there $\bar{B}_{g\odot}$ has also a small radial component; and 2) the three spacecrafts have different instantaneous distances from the Sun ($ \delta r / r \lesssim 0.6\%$) and therefore experience a different strength of the dipolar field. In the following, we shall neglect, for sake of simplicity, both corrections.
$\eta$ is the angle between the normal to the plane of the triangle and the axis of the Sun. Since the latter is in turn inclined with respect to the north of the ecliptic  \cite{SunJ} by the angle $\chi\simeq 7^\circ$, $\eta$ changes periodically during the year oscillating between $\eta_{min}=\beta-\chi$ and $\eta_{max}=\beta+\chi$; $\beta$ and $\chi$ are shown in Fig (\ref{fig1}b).

\section{ Proposed measurements}
The experiment we would like to consider uses LISA as the triangular closed path along which  e.m. signals are sent in opposite directions, in order to measure the ToF difference. Letting, for the moment, all practical problems aside, we note a multiplicity of contributions simultaneously acting upon the interferometer. GR is a non-linear theory so that, in principle, it is not an easy task to disentangle the various terms. In particular, beside the solar and galactic contribution, we have the kinematic terms related to the rotational movements of the configuration, i.e. the proper rotation of the triangle, its orbital motion around the Sun and the rotation of the Sun about the axis of the Milky Way. These kinematic terms may legitimately be considered as manifestations of the Sagnac effect; in a treatment that wanted to be "exact" all terms (both kinematic signals and "physical" i.e. GM terms) combine non-linearly with each other, so that it is extremely difficult to distinguish them. Fortunately, the fact that all effects are small or very small (the GM terms are expected to be much weaker than the kinematic ones) helps us, in the sense that in an approximate treatment we are allowed to truncate the expansion at the lowest order. Indeed the lowest approximation is the linear one where a simple superposition principle is applied.

Even accepting the linear approximation, the problem remains of separating the different addends of the sum. This separation would be possible if we knew at least the kinematic terms with an accuracy better than the size of the unknown "physical" contributions.
This is a difficult task, because the gravitomagnetic effect is expected to be many orders of magnitude smaller than the Sagnac effect; we need some sort of signature that highlights the terms of interest  and the "marking" we may exploit is the time dependence of the signals. Looking at Fig.(\ref{fig1}b) we see that our triangle oscillates yearly with respect to the galactic plane between an angle $\gamma_{max}\simeq\alpha+\beta$ and an angle $\gamma_{min}\simeq\alpha-\beta$,
where $\alpha \simeq 60^o$ is the angle between the ecliptic and galactic planes. The oscillation is not simply sinusoidal and we use $\simeq$ instead of just $=$ because the axes of the Milky Way, of the ecliptic and of the LISA triangle never turn out to be exactly co-planar.
Furthermore, when looking at the solar contribution, we find the already mentioned oscillation with respect to the solar spin axis.
\begin{figure}
    \centering
    \includegraphics[width=\textwidth]{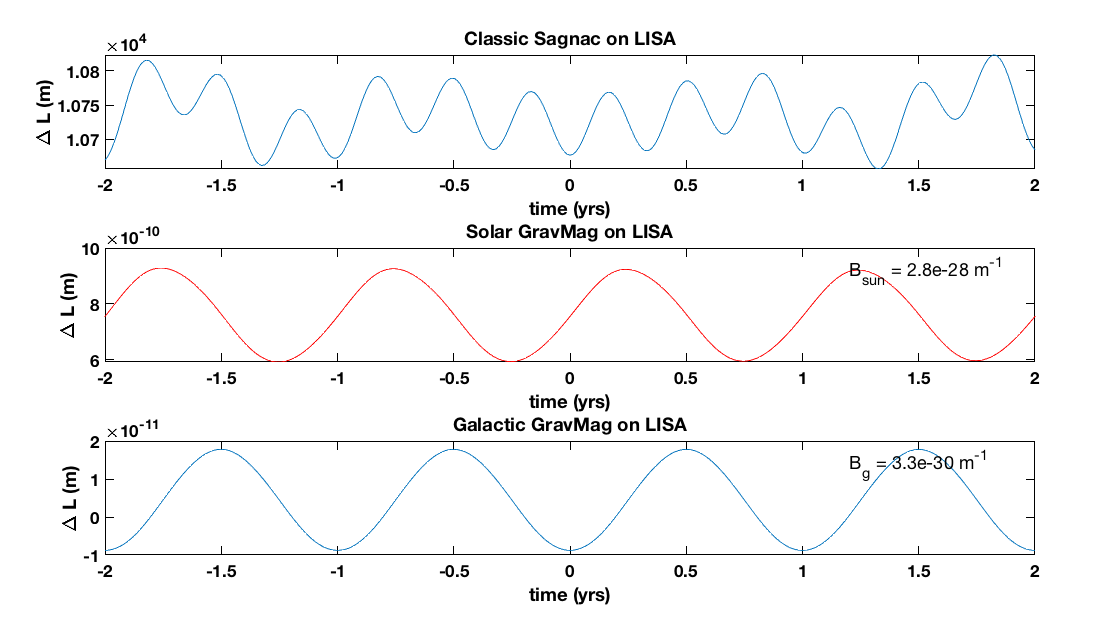}
    \caption{Time dependence of the expected signals (expressed as lengths) on a noiseless LISA constellation: \\
    TOP: Sagnac effect, due to the triangle rotations both around the Sun and around its center.  The signal has an almost constant value, with small ($\sim 1\%$) oscillations due to the flexing of the arms and ensuing change of the area. \\
    MIDDLE: GM effect from the Sun; the yearly modulation is due to the inclination of $J_\odot$ with respect to the ecliptic. \\
    BOTTOM:  signal from the galactic Gravito Magnetism; this measurement would be performed {\it within} the source mass current, and no certainties can be offered about its strength. We hypothized, for this plot, a field $B_{MW} = 3\cdot 10^{-30} m^{-1}$, neglecting the contribution of the external part of the Milky Way (that would actually act to reduce this strength).
}
    \label{fig:3plots}
\end{figure}

Thus, both GM contributions exhibit yearly periodicity but with a different phase, whereas the Sagnac effect is nearly constant with small harmonic components (including the yearly one) due to flexing, which  can be, in principle, calculated and subtracted.

As for the detectable signal, we may estimate its minimal amplitude for the Sun GM from eq. (\ref{Sundelay}), with a LISA constellation area, neglecting the above mentioned oscillations, $S=(\sqrt{3}/4)L^2\simeq2.7\times10^{18}$m$^2$.
The target displacement sensitivity for LISA,
is   $c |\Delta \tau_{min}|\sim10^{-11}$ m.
With this sensitivity both the solar GM and the galactic one should be detectable by LISA (see fig.\ref{fig:3plots}), although with marginal signal to noise ratio. \\ Eq. \ref{base} shows that the minimum detectable GM field would then be

\begin{equation}
B_{(min)}\sim 1.8\times10^{-30} \text{m}^{-1}\label{sens}
\end{equation}

In order to see what this number means, consider that the terrestrial GM field on the surface of the Earth (at the equator) is $\sim 10^{-22}$m$^{-1}$. Proper Sagnac terms (which depend on the angular velocities around the Sun, around the constellation center and around the galactic center) are many orders of magnitude larger than the value in eq.(\ref{sens}), whereas the solar contribution  is only a factor $100$  above the minimum detectable signal. On the basis of estimates of the angular momentum of the visible mass,  the galactic GM field is expected to be just above the edge of detectability.

\section{GM effect on light polarization}
The only GM effect we have considered so far is the asymmetry in the ToF of light, but there are other consequences to be considered. The analogy with classical e.m. suggests that we may also expect  the analog of the Faraday effect. In other words, if a linearly polarized light beam propagates along the field lines of a GM field the direction of the polarization vector should rotate around the direction of the propagation. \cite{ruggiero2007}

Using the formalism of GR, the electromagnetic field is described (in four dimensions) by the so called Faraday anti-symmetric tensor $F_{\mu\nu}$ and, regardless of any approximation, when there is propagation producing a four-dimensional area $\delta S^{0i}=(\delta\tau \delta x^i)=(\delta x^i)^2/c$  (movement along the direction of $x^i$ at the speed of light) the corresponding change in $F_{\mu\nu}$ is
\begin{equation}
\delta F^{\mu\nu}=(R^{\mu}_{\epsilon0i}F^{\epsilon \nu}+R^{\nu}_{\epsilon0i}F^{\mu\epsilon})\delta S^{0i}
\label{Riem}
\end{equation}
$R^{\mu}_{\nu\lambda\rho}$ are the components of the Riemann tensor which account for the curvature of space-time.

Whenever the propagation takes place over a distance $l$ in a direction non-perpendicular to the GM field, the change in the Faraday tensor corresponds to a rotation of the polarization vector of the wave around the propagation line by an angle $\psi$:\cite{ruggiero2007}

\begin{equation}
\psi\simeq -\frac{\sqrt{g_{00}}}{2}\bar{B}_g\cdot\bar l=\pm \frac{l}{2}\frac{N}{\sqrt{U}}(\frac{\partial_rN}{N}-\frac{\partial_rU}{U})
\label{faraday}
\end{equation}

The last expression in Eq. (\ref{faraday}) refers to the case of the Milky Way and uses the notation introduced in Eq. (\ref{bg}) applied on the galactic plane and for light travelling in a direction perpendicular to that plane.

 The solar GM will produce, in any  LISA arm, a rotation of the polarization plane of $ \sim 3 \cdot 10^{-19}$ prad.  The two laser beams propagating in each arm in opposite directions would undergo opposite rotations. In principle the effect would cumulate with the laser beam going around the constellation if the transmission of the light from satellite to satellite preserves the polarization.
LISA will use linearly polarized light for the strain measurements. The unwanted polarisation component  present in the received beam will be removed, when entering the measurement chain, by a PBS (polarising beam splitter);  this fraction, containing the GM signal, could be in principle detected by an additional, dedicated photo diode.
In fact, in the LISA-Pathfinder (LPF) mission it has been observed that, because of non perfect performance of the PBS, part of this light can still contaminate the strain measurement, yielding additional noise in the form of read-out noise and spurious laser pressure. Thus the GM rotation effect  in LISA would  possibly result in a decreased sensitivity of the observatory. That signal could be reconstructed in post-processing, as it was done in LPF, using a well shaped model of the measurement chain. \cite{Brig}

\section{Conclusion}
We have discussed the relevant sources that could contribute to the gravito-magnetic field in the solar system far from the planets and within 1 AU from the Sun. In particular we have considered the relevance of the field in producing an asymmetry in the propagation of electromagnetic signals along closed space paths. We have then discussed the possibility to use the big triangle formed by the LISA satellite constellation applying an approach similar to that of the Sagnac effect.

A quantitative evaluation of the effects shows that the sensitivity of such an arrangement would be sufficient to measure the solar GM field. Such measurement would give an important information on the internal structure of the Sun, from which the angular momentum of our star depends. But LISA, used $\grave{a}$ {\em  la Sagnac}, would probably be able to reveal the weak GM field of the Milky Way as well. This is increasingly interesting and important since the galactic gravitomagnetism depends of course on the distribution of visible matter in the Milky Way, but also,in a non marginal way, on the presence of dark matter, that, on the whole, is expected to have a mass much greater than that visible.

The GM field also causes a rotation of the polarization of the e.m. signals that will travel along the arms of the interferometer.  The effect is tiny , and its detection is extremely challenging, but should in principle be considered, in view of future technological advances. This would yield additional and complementary information on the intensity of the field.

LISA will send its ``raw" data to Earth, to be off-line processed and syntethized in interferometric  signals:  in view of this strategy, pursuing the search for GM signatures on the data, a challenging, though not impossible task,  should require no further hardware development, and no additional task for the mission.

For these reasons and for the importance of the objectives pursued it would be worthwhile to carry out the experiment.

\end{document}